\pdfoutput=1
\documentclass[a4paper,USenglish]{lipics}

\usepackage[lined,boxed,linesnumbered]{algorithm2e}
\SetAlgoInsideSkip{smallskip}
\usepackage{newfloat}

\DeclareFloatingEnvironment[fileext=lop,name=Algorithm]{algo}

\title{From Constrained Delaunay Triangulations to Roadmap Graphs with
Arbitrary Clearance}

\titlerunning{From CDTs to Roadmap Graphs with
Arbitrary Clearance}

\author{Stéphane Lens}
\author{Bernard Boigelot}
\affil{Montefiore Institute, B28\\
University of Liège\\
B-4000 Liège, Belgium\\
\texttt{\small\{lens,boigelot\}@montefiore.ulg.ac.be}}

\authorrunning{S. Lens and B. Boigelot}
\Copyright{Stéphane Lens and Bernard Boigelot}


\begin{document}

\maketitle

\begin{abstract}
This work studies path planning in two-dimensional space, in the
presence of polygonal obstacles. We specifically address the problem
of building a roadmap graph, that is, an abstract representation of
all the paths that can potentially be followed around a given set of
obstacles. Our solution consists in an original refinement algorithm
for constrained Delaunay triangulations, aimed at generating a roadmap
graph suited for planning paths with arbitrary clearance. In other
words, a minimum distance to the obstacles can be specified, and the
graph does not have to be recomputed if this distance is modified.
Compared to other solutions, our approach has the advantage of being
simpler, as well as significantly more efficient.
\end{abstract}

\section{Introduction}

Path planning is a central issue in fields such as mobile robotics;
autonomous robots often need to be able to compute on-the-fly
trajectories from their current position to a target location, taking
into account various feasibility and optimality constraints.

We study path planning in two-dimensional space in the presence of
obstacles. This problem has historically been addressed by various
approaches ranging from cell
decomposition~\cite{kambhampati1985multiresolution} to optimization,
potential field~\cite{ge2002dynamic}, and stochastic~\cite{lavalle01}
techniques. In this work, we address the construction of a
\emph{roadmap graph}, which is an abstract representation of all the
paths that can potentially be followed around the
obstacles~\cite{latombe1991planning,lavalle2006planning}. After it has
been computed, a roadmap graph can be explored using algorithms such
as Dijkstra's or A$^*$, in order to extract its shortest path from an
origin to a target destination, according to the metric of
interest. In actual applications, this path then needs to be
interpolated into a feasible trajectory, taking into account physical
constraints. In addition to robotics, roadmap graphs are also
especially useful in the case of problems that require to plan a large
number of paths constrained by the same set of obstacles, such as
crowd path planning~\cite{ali2013modeling}.

This article tackles the problem of building a roadmap graph for a
region of the Euclidean plane constrained by a given finite set of
polygonal obstacles. In addition to the geometrical description of
these obstacles, we also consider a \emph{clearance distance} $c$ that
defines the smallest distance at which the obstacles can be
approached. This is equivalent to studying the path planning
problem for a disk of radius $c$ that moves between obstacles
without intersecting them.

This problem has already been well studied. In the particular case of
obstacles taking the form of individual points, a roadmap graph can be
obtained by first computing a \emph{Voronoi diagram} of these
points~\cite{berg2008computational}. This yields a finite graph whose
nodes correspond to the points of the plane that locally maximize the
distance to their nearest obstacles. The edges of the graph represent
direct paths between nodes that clear the obstacles at the largest
possible distance, which can then be checked against $c$.

Thanks to a duality argument, the exact same roadmap graph can also be
constructed starting from a \emph{Delaunay triangulation} of the set
of points~\cite{devillers1998improved}.  The nodes of the graph are
then associated to the triangles of the triangulation (technically, to
the center of their circumcircle), and two nodes are linked by an edge
whenever their corresponding triangles share a common side. An
interesting property of such roadmap graphs is that they can be
constructed independently from the value of the clearance distance
$c$: Exploring the graph for a particular value of $c$ simply amounts
to following only the edges that correspond to triangle sides with a
length at least equal to $2c$. This property is of particular
importance to autonomous robotics applications: After having computed
a roadmap graph, evaluating several strategies for clearing the
obstacles at different margins of safety becomes possible at very
small cost.

Most applications require however to build roadmap graphs for sets of
obstacles that include line segments in addition to individual points,
which is the case in particular whenever obstacles take
polygonal shapes.  A first possible solution to this generalized
problem consists in computing a generalized Voronoi diagram, which
partitions the points of the plane according to their nearest point or
line obstacle~\cite{geraerts2010planning}. The main drawback is that
the edges of such diagrams can take the form of parabolic arcs, that
are more difficult to work with than line segments. A workaround consists
in discretizing line obstacles into finitely many individual
points~\cite{bhattacharya2008roadmap}, but it then becomes difficult
to find a good trade-off between a fine discretization level, which
may yield an unnecessarily large graph, and a coarse one, which might
result in an imprecise approximation. A third strategy is to build a
\emph{visibility-Voronoi complex}~\cite{wein2005visibility}, which
produces a graph from which a roadmap can easily be extracted, but
this technique turns out to be prohibitively costly, requiring a
computation time that is more than quadratic in the number of
obstacles.

A fourth solution is to start from a \emph{constrained Delaunay
  triangulation}, in which the line obstacles are required to appear
as sides of the triangles~\cite{chew1989constrained}. A roadmap graph
can then be extracted from such a triangulation using the same method
as for (classical) Delaunay ones.  Unfortunately, exploiting such a
graph for planning paths with a given clearance distance $c$ is not
as straightforward as before: The existence of a path in the graph
from a triangle to another that only traverses sides of length at
least equal to $2c$, does not guarantee anymore that there exists a
corresponding path in the plane, staying at a distance greater than
or equal to $c$ from the obstacles. In~\cite{kallmann2014dynamic},
this problem is solved by augmenting the triangulation with additional
parameters that have to be taken into account when the roadmap is
searched for paths. In addition, some parts of the triangulation have
to be refined in order for this mechanism to behave correctly.

The contribution of this work is to show that a roadmap graph suited
for arbitrary values of the clearance distance $c$ can be derived
from a constrained Delaunay triangulation of the obstacles. We
introduce a refinement algorithm for such triangulations, that
proceeds by inserting a carefully selected set of additional point
obstacles known as \emph{Steiner points}.  The resulting graph
precisely describes the possible paths in the plane around the
original set of obstacles. It has the property that the graph paths
traversing triangle sides of length at least equal to $2c$ exactly
correspond to the paths in the plane that stay at distance greater
than or equal to $c$ from the obstacles, for arbitrary values of $c$.
The method does not require to store additional information in the
triangulation, i.e., only the length of traversed triangle sides has
to be checked, which makes searching for paths very efficient. Our
algorithm has been implemented and experimentally evaluated on
randomly generated sets of obstacles, obtaining significantly lower
execution times than~\cite{bhattacharya2008roadmap,geraerts2010planning,kallmann2014dynamic,wein2005visibility}.

The paper is structured as follows. In Section~\ref{sec-points}, we
discuss the construction of a roadmap graph in the case of point
obstacles, using a Delaunay triangulation. In
Section~\ref{sec-points-lines}, we generalize this result to obstacles
composed of line segments in addition to individual points. Our
algorithm for refining constrained Delaunay triangulations is
developed in Section~\ref{sec-algorithm}.  Finally, we present in
Section~\ref{sec-experiments} experimental results, and
discuss in Section~\ref{sec-conclusions} the benefits of our approach.

\section{Point Obstacles}
\label{sec-points}

In this section, we recall classical results related to the
construction of a roadmap graph with respect to point obstacles.
Consider a finite set $\mathit{Obst} = \{A_1, A_2, \ldots, A_p \}$ of
points representing obstacles that must be avoided. A
\emph{triangulation} of these points is a finite set $\{ T_1, T_2,
\ldots, T_q \}$ of triangles taking their vertices in $\mathit{Obst}$,
such that their union exactly covers the convex hull of $A_1$, $A_2$,
\ldots $A_p$, and the intersection between any pair of triangles can
either be empty, equal to a single vertex $A_i$, or equal to a segment
$[A_i A_j]$ linking two points in $\mathit{Obst}$.  Such a
triangulation is said to be \emph{Delaunay} if, for every triangle
$T_i$, all the points in $\mathit{Obst}$ are located outside or on the
circumcircle of $T_i$~\cite{bern2004triangulations}.

From a Delaunay triangulation of $\mathit{Obst}$, one can extract a
roadmap graph representing the possible paths in the plane around
those obstacles. This is done by building a graph $G$ whose nodes
correspond to the triangles $T_i$, and in which two nodes are linked
by an edge if and only if their underlying triangles $T_i$ and $T_j$
share a common side.

An interesting property of this graph is that it can be used for
reasoning about the possible paths in the plane that remain at a
specified \emph{clearance distance} $c$ from obstacles, even though
the graph is defined independently from $c$. Let us define a
\emph{feasible position} $P$ as a point in the plane such that $|P -
A_i| \geq c$ for all $i$. A \emph{feasible path} is defined as a path
in the plane that only visits feasible positions. A \emph{feasible
  triangle} is one that contains at least one feasible point. The
following result is well 
known~\cite{bern2004triangulations,berg2008computational}.

\begin{theorem}
\label{theo-triangul-roadmap}
There exists a path in $G$ from a feasible triangle $T_i$ to another
one $T_j$ that traverses only triangle sides of length $\ell$
such that $\ell \geq 2c$ iff there exists a feasible path in the plane leading
from a point in $T_i$ to a point in $T_j$, with respect to the
clearance distance $c$.
\end{theorem}


This theorem intuitively expresses that the feasible paths in the
plane that can be followed around the obstacles are exactly
represented by the paths in the graph $G$, being careful of only traversing
triangle sides that have a length consistent with the clearance distance. The
graph $G$ thus represents a roadmap that can be searched for paths
leading from a triangle to another, using for instance algorithms such
as Dijkstra's or A$^*$~\cite{wagner2007speed}.

It is worth mentioning that a path $\pi$ of $G$ that visits a sequence
of triangles $T_{i_1}, T_{i_2}, T_{i_3}, \ldots$ does not always
translate into a feasible path in the plane that exactly follows this
sequence of triangles. It may indeed be the case that moving from
$T_{i_k}$ to $T_{i_{k+1}}$ requires to pass through an intermediate
triangle that is not represented in the path $\pi$, which is not
problematic. However, the sequence of triangles successively visited
by any feasible path in the plane must necessarily correspond to a
path that exists in the graph $G$.

After having extracted a path $\pi = T_{i_1}, T_{i_2}, \ldots, T_{i_m}$ from
$G$, for a given clearance distance $c$, a corresponding feasible path in
the plane can be obtained as follows. The goal is to generate a path
that clears the obstacles in the same way as $\pi$, but that is
locally optimal in the sense that it minimizes the traveled
distance and does not contain unnecessary switchbacks.
The sequence of triangles $T_{i_1}, T_{i_2}, \ldots, T_{i_m}$ represents a
\emph{channel}, i.e., a triangulated region of the plane in which one 
can move from a triangle $T_{i_k}$ to its successor $T_{i_{k+1}}$ by
traversing their common side. The vertices of the triangles that
compose a channel precisely represent the obstacles that must be
cleared when moving along this channel.

Given a channel obtained for a clearance distance $c$, the shortest
feasible path in the plane that follows this channel can be computed
thanks to a simple generalization of the \emph{funnel
  algorithm}~\cite{demyen2006efficient}. This operation conceptually
consists in placing a solid disk of radius $c$ on each channel vertex,
and of pulling an elastic cord taunt between these disks, from the
entrance to the exit triangles of the channel.

The main advantage of this approach to synthesizing paths is its
efficiency: Triangulating a set of points and searching for the
shortest path in a finite graph can be performed in $O(n \log n)$
time, where $n$ denotes the number of obstacles, and the funnel
algorithm runs in $O(n)$ time. (The practical cost of this algorithm
is usually even smaller, since it actually runs in linear time with
respect to the number of triangles that belong to the considered
channel.)  One drawback is the fact that the shortest path in the
roadmap graph does not necessarily correspond to the shortest feasible
one in the plane, but using a suitable metric, it is known that this
approximation usually suffices for most
applications~\cite{kallmann2010shortest}.

\section{Point and Line Obstacles}
\label{sec-points-lines}

Our aim is now to generalize the results of Section~\ref{sec-points}
to sets of obstacles that include line segments in addition to
individual points. The motivation is to be able to deal with obstacles
that have a polygonal shape. We first adapt triangulations to this
setting, and then investigate how to build roadmap graphs out of them.

\subsection{Constrained Triangulations}

Let us consider a set of obstacles $\mathit{Obst}= \{ A_1, A_2, \ldots
A_p, L_1, L_2, \ldots, L_q \}$ composed of a finite number of
points $A_i$ as well as a finite set of line segments $L_i$. We assume
w.l.o.g. that each segment links two points $A_i$ and $A_j$ that
belong to $\mathit{Obst}$, and that the intersection of any pair of
segments is either empty, or limited to a point $A_i$ belonging to
$\mathit{Obst}$. It is also natural to require that path planning
remains restricted to areas that are fully delineated by external obstacles.

\begin{figure}
\centerline{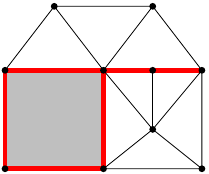}
\caption{Example of constrained triangulation.}
\label{fig-ex-constr-triangul}
\end{figure}

For a set of obstacles $\mathit{Obst}$, a \emph{constrained
  triangulation} is a triangulation in which every segment $L_i \in
\mathit{Obst}$ (called a \emph{constrained segment}) forms the side of
at least one triangle. Note that, if the goal is to reason about
feasible paths in the plane, the notion of triangulation can somehow
be slightly extended. First, it is not mandatory for regions that
cannot be reached, such as the interior of polygonal obstacles, to be
covered by triangles. Second, since constrained segments cannot be
traversed, we allow a side of a triangle to be composed of a series of
collinear constrained segments instead of a single one. An example of
constrained triangulation is given in
Figure~\ref{fig-ex-constr-triangul}, in which an unreachable region is
grayed out, and constrained segments are represented in red.

Delaunay's criterion can readily be adapted to constrained
triangulations.  Given a set $\mathit{Obst}$ of obstacles, a
point $P_2$ is said to be \emph{visible} from a point $P_1$ 
if the intersection between
the segment $[P_1P_2]$ and every constrained segment $[A_i A_j] \in
\mathit{Obst}$ is either empty, or is equal to one of its extremities
$P_1$ or $P_2$. A constrained triangulation is then \emph{Delaunay} if
for each of its triangles $A_i A_j A_k$, all the points in
$\mathit{Obst}$ that are visible from $A_i$, $A_j$ or $A_k$ are
located outside or on its circumcircle. It is easily shown that this
criterion is equivalent to imposing that, for every pair $(A_i A_j
A_k, A_{\ell} A_j A_i)$ of adjacent triangles, with $A_k \neq A_{\ell}$,
either $[A_i A_j]$ is a constrained segment, or one has
$\widehat{A_iA_kA_j} + \widehat{A_iA_{\ell}A_j} \leq \pi$.

\subsection{Refining Constrained Triangulations}

\begin{figure}
\centerline{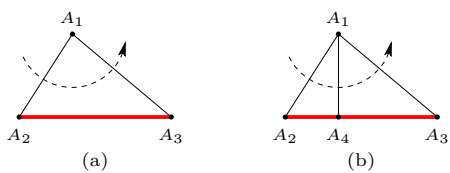}
\caption{Constrained triangulation refinement.}
\label{fig-constr-segment}
\end{figure}

A roadmap graph can be derived from a constrained triangulation of a
set of obstacles by a procedure similar to the one outlined in
Section~\ref{sec-points}: One builds a graph whose nodes are
associated to the triangles, and links the nodes corresponding to pairs of
adjacent triangles. However, in the presence of constrained segments,
Theorem~\ref{theo-triangul-roadmap} does not hold anymore.  The
problem is illustrated in Figure~\ref{fig-constr-segment}(a), where the
clearance distance $c = \frac{1}{2}\min(|A_1 A_2|, |A_1 A_3|)$ does not
make it possible to
traverse the gap between the point $A_1$ and the constrained
segment $[A_2 A_3]$.

Our solution to this problem is to \emph{refine} the constrained
triangulation by inserting additional obstacle points, known as
\emph{Steiner points}, until we obtain a constrained triangulation for
which a result similar to Theorem~\ref{theo-triangul-roadmap} can be
established.  Adding a Steiner point amounts to splitting a
constrained segment in two components connected at that point,
fragmenting one of the two triangles adjacent to this segment. For
instance, Figure~\ref{fig-constr-segment}(b) shows that adding the
orthogonal projection $A_4$ of $A_1$ onto $[A_2 A_3]$ as a Steiner
point decomposes the triangle $A_1 A_2 A_3$ into $\{ A_1 A_2 A_4, A_1
A_4 A_3 \}$, yielding a channel for which
Theorem~\ref{theo-triangul-roadmap} now applies. Indeed, it now becomes
possible to traverse the sequence of segments $[A_1 A_2]; [A_1 A_4] ; [A_1
  A_3]$ if and only if the clearance distance $c$ is at most equal to $
\frac{1}{2}\min(|A_1 A_2|, |A_1 A_4|, |A_1 A_3|) = \frac{1}{2} |A_1 A_4|$.

Since inserting a Steiner point splits a triangle in two, this
operation may result in a refined constrained triangulation that does
not satisfy anymore Delaunay's criterion. It then becomes necessary to
update the triangulation so as to restore this property, which will be
essential to the correct computation of further Steiner points. This update
procedure is carried out as follows. Consider a triangle $A_1 A_2
A_3$ that has been split into $\{ A_1 A_2 A_4, A_1 A_4 A_3 \}$ by the
addition of a Steiner point $A_4$ inside the segment $[A_2 A_3]$. If
the segment $[A_1 A_2]$ is unconstrained and joins two triangles $A_1
A_2 A_4$ and $A_1 A_5 A_2$, then one has to check whether they satisfy
Delaunay's criterion. If not, the quadrilateral $A_4 A_1 A_5A_2$ must
be \emph{flipped}, which consists in replacing the pair of triangles
$\{ A_1 A_2 A_4, A_1 A_5 A_2 \}$ by $\{ A_1 A_5 A_4, A_2 A_4 A_5
\}$. The same operation must be performed for the segment $[A_1 A_3]$,
as well as for the remaining outside edges of flipped
quadrilaterals. It is known that this procedure, borrowed from
\emph{incremental Delaunay triangulation} algorithms, has a low
average cost, the expected number of triangles to be processed
remaining bounded~\cite{berg2008computational}.

Finally, it should be stressed out that the refinement operation must
be able to handle situations that are more complex than the one
illustrated in Figure~\ref{fig-constr-segment}.  For instance, even in
situations where $[A_2 A_3]$ is unconstrained, the traversal of $[A_1
  A_2]; [A_1 A_3]$ might be affected by constraints located beyond
this segment. We address the problem of selecting a correct set of
Steiner points in the next two sections.

\subsection{Point/Point and Point/Line Constraints}

Recall that the goal is to obtain a roadmap graph in which, for every
clearance distance $c > 0$, each path that only traverses triangle
sides of length $\ell$ such that $\ell \geq 2c$ represents a channel
that can be passed by remaining at a distance at least equal to $c$
from the obstacles.  In other words, one should be able to check
whether a channel can be passed with sufficient clearance only by
measuring the length of the traversed triangle sides, similarly to the
case of point obstacles discussed in Section~\ref{sec-points}.

A solid disk of radius $c$ that moves along a channel has to clear
different configurations of obstacles. First, traversing a segment $[A_m
  A_n]$ between two vertices $A_m$ and $A_n$ is only possible if $c
\leq \frac{1}{2}|A_m A_n|$. We call this condition a \emph{point/point
  constraint}. Second, if the orthogonal projection $H$ of a vertex
$A_k$ on a constrained segment $[A_i A_j]$ is located inside this
segment, moving between $A_k$ and this segment (in other words,
traversing $[A_k H]$) imposes $c \leq \frac{1}{2}|A_k H|$, which forms
a \emph{point/line constraint}.  In order to assess the possibility of
passing a channel for a given clearance distance $c$, it is sufficient
to consider point/point and point/line constraints.  As discussed
in~\cite{kallmann2014dynamic}, other configurations such as moving
between two constrained segments are systematically covered by the
point/point and point/line constraints induced by their extremities.

It has been shown in Section~\ref{sec-points} that point/point
constraints are automatically dealt with by Delaunay triangulations:
In the absence of constrained segments,
Theorem~\ref{theo-triangul-roadmap} holds and the triangulation
provides a suitable roadmap graph. The difficulty is thus to handle
correctly point/line constraints.

Consider a triangle $A_1 A_2 A_3$ crossed from the side
$[A_1 A_2]$ to the side $[A_1 A_3]$, or the other way around.  We
assume w.l.o.g. that this triangle satisfies $|A_1 A_2| \leq |A_1
A_3|$. We check whether there exists a vertex $X$ of the triangulation
located on the same side of $A_2 A_3$ as $A_1$, that is involved in a
point/line constraint with a constrained segment $[A_i A_j]$ located on
the other side of this line. (This precisely means that the orthogonal
projection $X'$ of $X$ on $A_i A_j$ belongs to the interior of the
segment $[A_i A_j]$, and is such that $[XX']$ intersects $[A_2 A_3]$.)

A point/line constraint induced by such a vertex $X$ is relevant only
if it is more restrictive than the point/point constraints generated
by the channel that is followed. We introduce the following definition.

\begin{definition}
\label{def-problematic}
A vertex
$X$ is said to be \emph{problematic} for the traversed segments $[A_1 A_2]$ and
$[A_1A_3]$ if
\begin{itemize}
\item it is located on the same side of $A_2 A_3$ as $A_1$,
\item there exists a constrained segment $[A_i A_j]$ that contains
the orthogonal projection $X'$ of $X$, which is such that 
$[XX'] \cap [A_2 A_3] \neq \emptyset$, and
\item
the points $X$ and $X'$ satisfy $|XX'|< \min(|A_1A_2|, |XA_2|, |XA_3|)$.
\end{itemize}
\end{definition}

\begin{figure}
\centerline{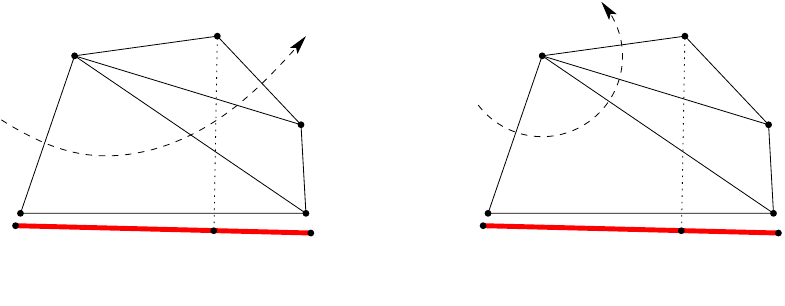}
\caption{Channel types.}
\label{fig-channel-types}
\end{figure}

Indeed, it is not necessary to consider vertices $X$ for which $|XX'|
\geq |A_1 A_2|$, since the point/line constraint between $X$ and $[A_i
  A_j]$ is then systematically satisfied as a consequence of the fact
that the segment $[A_1 A_2]$ can be traversed with sufficient
clearance.  The motivation for the conditions $|XX'| < |XA_2|$ and
$|XX'| < |XA_3|$ is less obvious.  With respect to $X$, there are
actually two types of possible channels in which the triangle $A_1 A_2 A_3$ is
traversed from $[A_1 A_2]$ to $[A_1 A_3]$: Those in which the segment $[XX']$
needs to be crossed (\emph{type~1}), and those in which
it does not (\emph{type~2}). Note that the segment $[XX']$ does not
necessarily correspond to the side of a triangle in the triangulation. The two
situations are illustrated in Figure~\ref{fig-channel-types}.

For a channel of type~1, the point/line constraint between $X$ and
$[A_i A_j]$ cannot be more restrictive than the point/point
constraints induced by the channel if the condition $|XX'| >
\min(|XA_2|, |XA_3|)$ is satisfied, since the latter guarantee that
the segment $[XX']$ can be traversed. On the other hand,
in a channel of type~2, the point/line constraint induced by $X$ is
irrelevant and does not need to be taken into account.

When studying the triangle $A_1 A_2 A_3$, one does not know whether
the channel that will be followed after traversing $[A_1 A_2]$ and
$[A_1 A_3]$ is of type~1 or~2, hence both possibilities have to be
taken into account. This goal is achieved by considering only the
point/line constraints induced by problematic vertices, according to
Definition~\ref{def-problematic}. Note that the notion of problematic
vertices captured by this definition is similar to the concept of
\emph{disturbances} introduced in~\cite{kallmann2010shortest}, but our
strategy for dealing with them, described in the next section,
completely differs.

\subsection{Steiner Points Placement}
\label{sec-steiner}

Our approach to selecting a suitable set of Steiner points is based on
the following idea: For every pair of unconstrained sides $([A_1A_2],
[A_1A_3])$ of every triangle $A_1A_2A_3$ of the constrained
triangulation, if this pair of segments potentially admits a
problematic vertex, then this triangle has to be refined. The
refinement operation consists in inserting a Steiner point that splits
the triangle in such a way that problematic vertices cannot exist
anymore.

Checking for problematic vertices is done as follows. Recall that we
assume w.l.o.g. $|A_1A_2| \leq |A_1A_3|$. First, it is easily
established that the pair $([A_1A_2], [A_1A_3])$ can only admit a
problematic vertex if the corresponding triangle is such that
$\widehat{A_1A_2A_3} < \pi/2$. This property relies on the fact that
the constrained triangulation satisfies Delaunay's criterion, from
which it follows that all the points located outside or on the
circumcircle of $A_1A_2A_3$ necessarily violate at least one of the
conditions to be problematic.

\begin{figure}
\centerline{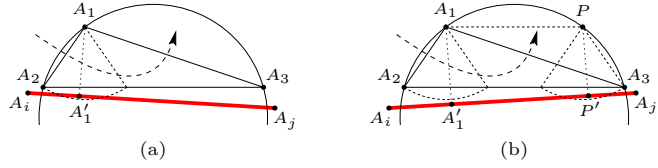}
\caption{Checking for problematic vertices.}
\label{fig-projection}
\end{figure}

If the angle $\widehat{A_1A_2A_3}$ is acute, which can be checked by
testing whether the orthogonal projection of $A_1$ onto $A_2A_3$ is an
interior point of the segment $[A_2A_3]$, then the next step is to check
whether $A_1$ itself is problematic. This is equivalent to deciding
whether there exists a constrained segment $[A_i A_j]$ that contains
as an interior point the orthogonal projection $A'_1$ of $A_1$ onto
$A_iA_j$, such that $A_1'$ is visible from $A_1$, $[A_1 A_1']$ crosses
$[A_2 A_3]$, and $|A_1 A_1'| < |A_1 A_2|$. We will introduce a
procedure for carrying out efficiently this check in
Section~\ref{sec-algorithm}. If this operation succeeds, then $A'_1$
is inserted as a Steiner point into the constrained segment $[A_i
  A_j]$, splitting the triangle that is adjacent to this segment, on
the same side as $A_1$. The situation is illustrated in
Figure~\ref{fig-projection}(a).

When, on the other hand, $A_1$ turns out to be unproblematic, it
remains to check whether there may exist other potential problematic
vertices. This operation is performed in the following way. We compute
the point $P$ located at the intersection of the circumcircle of
$A_1A_2A_3$ and the line parallel to $A_2 A_3$ containing $A_1$. Then,
we apply to $P$ the same decision procedure as to $A_1$ in the
previous step: We check whether $P$ projects onto a constrained
segment $[A_i A_j]$, its projection $P'$ is visible from $P$, $[PP']$
crosses $[A_2 A_3]$, and $|PP'| < |A_1 A_2|$. In such a case, the
projection $A'_1$ of $A_1$ onto $[A_i A_j]$ is necessarily interior to
this segment. We then add $A'_1$ as a Steiner point, in the same way
as before. This case is illustrated in
Figure~\ref{fig-projection}(b). The motivation for selecting $A'_1$
instead of $P'$ is twofold. First, $P$ is generally not a vertex of
the constrained triangulation, hence it does not always appear in the
result of the refinement. Second, the goal of getting rid of triangles
that may admit problematic vertices is correctly achieved with the
choice of $A'_1$.

\begin{figure}
\centerline{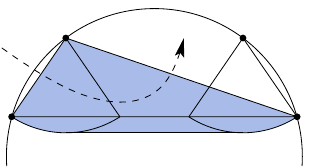}
\caption{Safe zone.}
\label{fig-safe-zone}
\end{figure}

Indeed, after having refined a constrained triangulation using the
procedure described in this section, every resulting triangle that is
part of a channel includes a \emph{safe zone} that cannot be crossed
by constrained segments, and allows the passage of a solid disk of
diameter equal to the smallest triangle side that is traversed. This
safe zone is illustrated in Figure~\ref{fig-safe-zone} for the
triangle considered in Figure~\ref{fig-projection}. As a consequence,
if a disk of given diameter is able to pass all the segments that
compose a channel, then it can traverse this channel while remaining
confined in the safe zone of the successively visited triangles. This
proves the following result.

\begin{theorem}
\label{theo-constr-triangul-roadmap}
In a refined constrained triangulation, there exists a path in $G$
from a feasible triangle $T_i$ to another one $T_j$ that traverses only
unconstrained triangle sides of length $\ell$ such that $\ell \geq 2c$ iff there exists a
feasible path leading from a point in $T_i$ to a point in $T_j$, with
respect to the clearance distance $c$.
\end{theorem}

Finally, it is worth mentioning that our refinement procedure always
terminates, since the insertion of a Steiner point creates right
triangles that will not be refined further.

\section{Refinement Algorithm}
\label{sec-algorithm}

As explained in Section~\ref{sec-steiner}, a constrained triangulation
is refined by checking, for each pair $([A_1A_2], [A_1A_3])$ of
unconstrained sides belonging to one of its triangles $A_1A_2A_3$,
whether a Steiner point has to be created by projecting $A_1$ on some
constrained segment. Assuming w.l.o.g. $|A_1A_2|\leq|A_1A_3|$, this check
only has to be performed if $\widehat{A_1A_2A_3} < \pi/2$, and amounts
to deciding whether there exists a constrained segment $[A_i A_j]$ onto
which $A_1$ can be projected, with its projection $A'_1$ satisfying
specific conditions. If the check does not succeed, a similar operation
has to be carried out for another point $P$ derived from $A_1$.

The crucial issue in the procedure is to perform efficiently the
check operation, which can be seen as a particular instance of the
following problem: Given a point $X$, a triangle side $[A_k A_{\ell}]$
in the triangulation that contains the orthogonal projection $X'$ of
$X$, and a distance $\lambda$ such that $|XX'| < \lambda \leq \min
(|XA_k|,|XA_{\ell}|)$, decide whether there exists a point $Y$
belonging to a constrained segment $[A_iA_j]$, such that $[XY]$
crosses $[A_kA_{\ell}]$ with $|XY|<\lambda$.  If the answer is
positive, return the corresponding segment $[A_iA_j]$ that is nearest
to $X$.

In order to solve this problem, the first step is to check whether
$[A_kA_{\ell}]$ itself is a constrained segment, in which case the
procedure directly terminates with $[A_iA_j]=[A_kA_{\ell}]$. Otherwise, we
explore the triangulation, starting with the triangle $A_kA_{\ell}A_q$
adjacent to $[A_kA_{\ell}]$ on the opposite side as $X$. It can be
shown that, as a consequence of Delaunay's criterion, the segment $[A_kA_q]$ can
contain a point $Y$ such that $|XY|<\lambda$ only if
$|A_kA_q|>|A_{\ell}A_q|$. Therefore, the search can be
continued by repeating the same procedure with the longest segment among
$[A_kA_q]$ and $[A_{\ell}A_q]$ replacing $[A_k A_{\ell}]$.  The check
terminates when either a suitable constrained segment is found, the
projection $X'$ of $X$ onto $[A_k A_{\ell}]$ does not exist, or this
projection is such that $|XX'|\geq \lambda$.

The complete algorithm for refining a constrained triangulation is
summarized in Algorithm~\ref{algo-refinement}.

\begin{algo}
\IncMargin{0.6em}
\begin{algorithm}[H]
  \small
  \SetKwFunction{Refine}{\footnotesize Refine}
  \SetKwFunction{Check}{\footnotesize Check}
  \SetKwFunction{FlipIfNeeded}{\footnotesize FlipIfNeeded}
  \SetKwBlock{myproc}{}{}
  \textbf{Function} \Refine{\rm constrained triangulation $\mathit{Tr}$}
  \myproc{
    \For{\rm all unconstrained $[A_1A_2]$, $[A_1A_3]$ such that
         $A_1A_2A_3 \in \mathit{Tr}$ and $|A_1A_2| \leq |A_1A_3|$}
    {
      $(\mathit{ok}, [A_iA_j])$ := 
        \Check{$\mathit{Tr}$, $A_1$, $[A_2A_3]$, $|A_1A_2|$}\;
      
      \If{\rm not $\mathit{ok}$}
      {
        $P$ := $\mathit{circumcircle}(A_1A_2A_3) \cap d$,
          with $d \parallel A_2A_3$, $A_1 \in d\:\!$\;
          
        $(\mathit{ok}, [A_iA_j])$ := 
          \Check{$\mathit{Tr}$, $P$, $[A_2A_3]$, $|A_1A_2|$}\;
      }
      
      \If{\rm $\mathit{ok}$}
      {
        $A'_1$ := \mbox{projection of $A_1$ onto $[A_iA_j]$}\;
        
        $A_k$ := \mbox{vertex of $A_iA_jA_k \in
         \mathit{Tr}$, on same side of $A_iA_j$ as $A_1$}\;
         
        $\mathit{Tr}$ := $(\mathit{Tr} \setminus
          \{ A_iA_jA_k \}) \cup \{ A_k A_i A'_1, A_k A'_1 A_j\}$\;
          
        \FlipIfNeeded{$\mathit{Tr}$, $A'_1$, $[A_kA_i]$}\;
        \FlipIfNeeded{$\mathit{Tr}$, $A'_1$, $[A_kA_j]$}\;
      }
    }
    
    \Return $\mathit{Tr\;\!}$\;
  }
  
  \BlankLine
  
  \textbf{Function} \Check{$\mathit{Tr}$, $X$, $[A_iA_j]$, $\lambda$}
  \myproc{
    $X'$ := \mbox{projection of $X$ onto interior of $[A_iA_j]$}\;
    
    \lIf{\rm $X'$ does not exist or $|XX'| \geq \lambda$}
        {\Return  $(\mathit{false}, -)\;\!$}
        
    \lIf{\rm $[A_iA_j]$ is constrained}
        {\Return  $(\mathit{true}, [A_iA_j])\;\!$}
        
    $A_k$ := \mbox{vertex of $A_iA_jA_k \in
      \mathit{Tr}$, on other side of $A_iA_j$ as $X$}\;
      
    \uIf{$|A_kA_i| > |A_kA_j|$}
    {
      \Return \Check{$\mathit{Tr}$, $X$, $[A_kA_i]$, $\lambda$}\;
    }
    \Else
    {
      \Return \Check{$\mathit{Tr}$, $X$, $[A_kA_j]$, $\lambda$}\;
    }
  }

  \BlankLine
  
  \textbf{Function} \FlipIfNeeded{$\mathit{Tr}$, $A_1$, $[A_2A_3]$}
  \myproc{
    \If{\rm $[A_2A_3]$ is unconstrained and adjacent to two triangles of
        $\mathit{Tr}$}
    {
      $A_4$ := vertex of $A_4A_2A_3 \in \mathit{Tr}$ such that
        $A_4 \neq A_1$\;

      \If{\rm $A_4$ is interior to $\mathit{circumcircle}(A_1A_2A_3)$}
      {
        \FlipIfNeeded{$\mathit{Tr}$, $A_1$, $[A_2A_4]$}\;
        \FlipIfNeeded{$\mathit{Tr}$, $A_1$, $[A_3A_4]$}\;
      }
    }
  }
\end{algorithm}
\DecMargin{0.6em}
\caption{Refinement algorithm.}
\label{algo-refinement}
\end{algo}

\section{Complexity and Experimental Results}
\label{sec-experiments}

The time needed to run our constrained Delaunay triangulation
refinement algorithm can be estimated as follows. Let $n$ denote the
number of obstacle vertices. The number of triangles that have to be
inspected at Line~2 of Algorithm~\ref{algo-refinement} is $O(n)$. Indeed, since we start
from a constrained Delaunay triangulation, the expected number of
triangles incident to vertices of the triangulation remains
bounded~\cite{berg2008computational}, hence the total number of
projections of vertices onto constrained segments performed by the
algorithm is $O(n)$. Each insertion of a Steiner point also entails a
bounded number of operations (Lines 12--13 and 27--34) in order to restore
Delaunay's criterion. The time needed to inspect one triangle (Lines
3--7) is however not bounded and can be as high as $O(n)$, which brings
the worst-case time complexity of the algorithm to $O(n^2)$.

We have been able to reach this worst-case upper bound with
specifically crafted families of instances, but only by
implementing the choice performed at Line~2 with a deliberately naive
approach, systematically selecting the triangle that leads to the
larger number of calls to \textit {check()}. With the more sensible
strategy of inspecting first the triangles that include exactly one
constrained side, which is easy to implement, we did not manage to
find instances of the problem for which the asymptotic cost of our
algorithm exceeds $O(n)$. The question of proving that the worst-case
complexity is actually linear using this simple heuristics remains
open, similarly to the solution proposed
in~\cite{kallmann2014dynamic}.

Our refinement algorithm has been implemented in order to compare it
experimentally to
solutions such as~\cite{geraerts2010planning,kallmann2014dynamic}. In
addition to being simpler, our technique also has the advantage of
offering significantly smaller execution times, even though it
produces slightly larger refined triangulations
than~\cite{kallmann2014dynamic}. We provide in
Figure~\ref{fig-runtimes} measurements performed on randomly generated
sets of obstacles of increasing size, using a i5-460M CPU at 2.53
GHz. Note that the reported values do not include the time needed
to compute the constrained Delaunay triangulation of the obstacles prior
to running our algorithm.

\begin{figure}
\captionsetup{aboveskip=1.5\normalbaselineskip}
\def\svgwidth{13cm}
\centerline{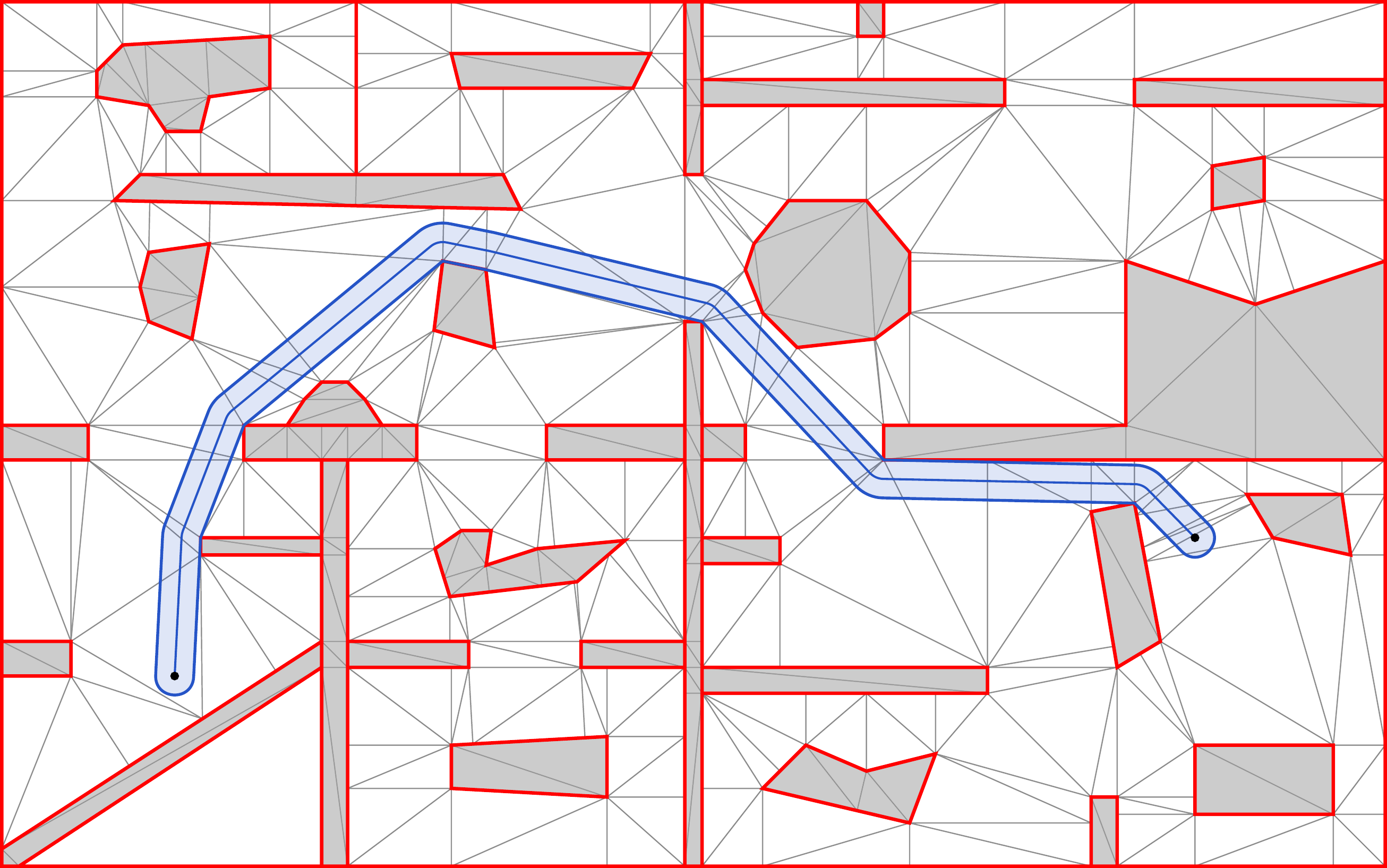}
\vspace{1.0cm}
\def\svgwidth{13cm}
\centerline{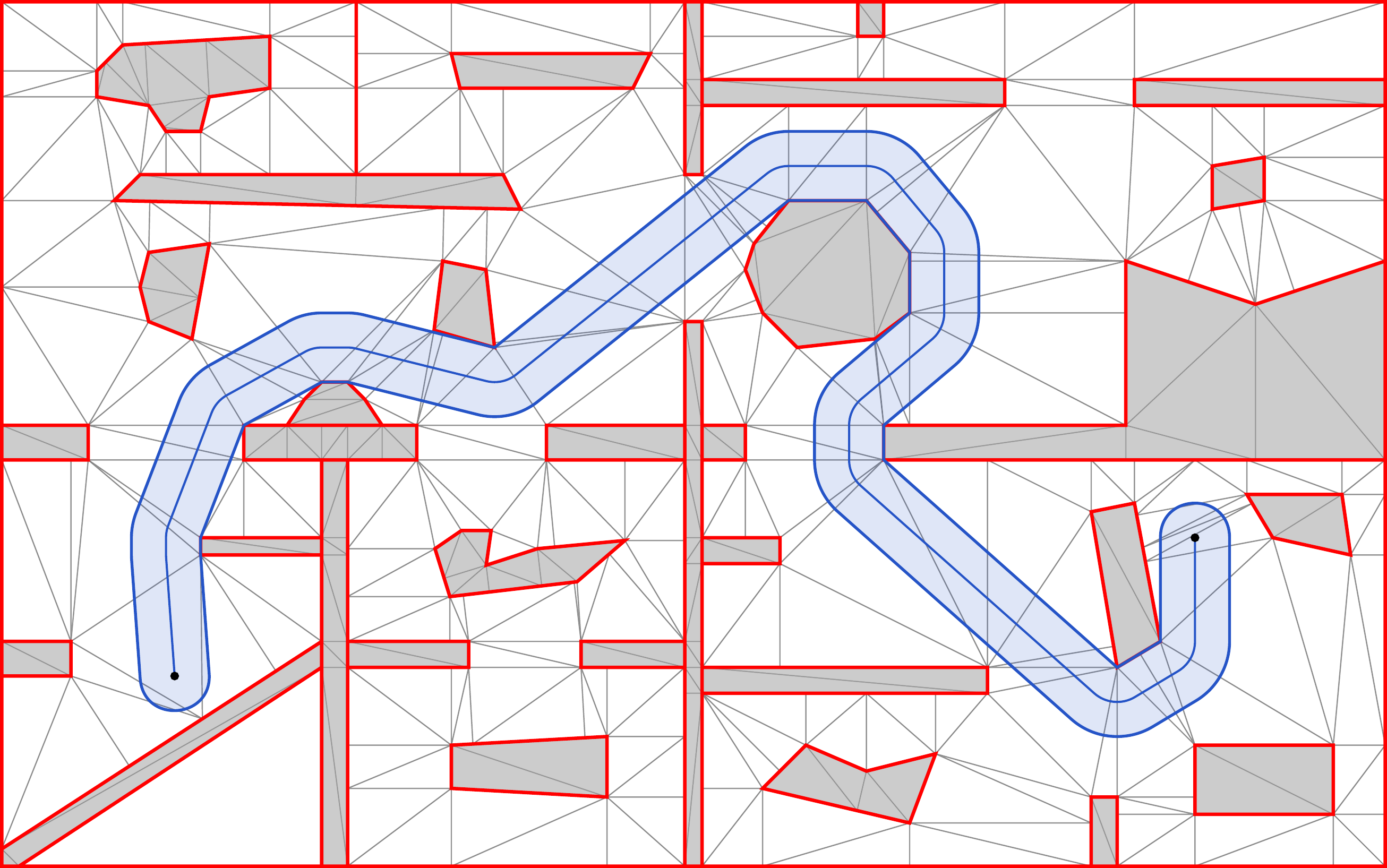}
\caption{Example of refined triangulation, and shortest paths for two
clearance distances.}
\label{fig-ex-roadmap}
\end{figure}

An example of refined constrained triangulation computed for a typical
case study is shown in Figure~\ref{fig-ex-roadmap}, with shortest
paths extracted for two different values of the clearance
parameter. For this example, the refinement procedure runs in about 30
$\mu$s.

\begin{figure}
\captionsetup{aboveskip=0.1\normalbaselineskip}
\begin{center}
\footnotesize
\begin{tabular}{rrrrr}
Points & Triangles & Refined points & Refined triangles & Time (ms)\\\hline
\\[-0.9em]
1227 & 2448 & 1420 & 2641 & 0.27\\
5558 & 11110 & 6179 & 11731 & 1.23\\
15042 & 30078 & 16644 & 31680 & 3.49\\
30205 & 60404 & 33311 & 63510 & 9.08\\
54990 & 109974 & 60563 & 115547 & 19.78\\
136168 & 272330 & 150022 & 286184 & 52.66\\
325058 & 650110 & 358931 & 683983 & 128.31\\
649632 & 1299258 & 719149 & 1368775 & 260.05\\
1298879 & 2597752 & 1443726 & 2742599 & 526.40
\end{tabular}
\end{center}
\caption{Experimental results.}
\label{fig-runtimes}
\end{figure}

\section{Conclusions}
\label{sec-conclusions}

In this work, we have introduced an original method for extracting a
roadmap graph from a constrained Delaunay triangulation of a set of
polygonal obstacles. Our approach consists in refining the constrained
triangulation by adding a carefully selected set of Steiner
points. The resulting triangulation has the important property that it
can be searched for paths in the plane that avoid the obstacles at an
arbitrarily selected \emph{clearance distance} $c$: this amounts to
checking that the triangle sides that are traversed have a length
$\ell$ that satisfies $\ell \geq 2c$. This contrasts with techniques
such as~\cite{kallmann2014dynamic} that require to store additional
information in the triangulation, and to modify search algorithms in
order to take it into account. Compared to the algorithm
proposed in~\cite{lamarche2004crowd}, ours is able to handle constraints
induced by a point and a segment that do not belong to the same
triangle, such as, e.g., the situations depicted in
Figure~\ref{fig-projection}.  Our method has been implemented, and has
the advantages of being simpler and significantly more efficient than
other known solutions such
as~\cite{bhattacharya2008roadmap,geraerts2010planning,kallmann2014dynamic,wein2005visibility}. Its
drawback is that it produces refined triangulations that are slightly
larger than those of~\cite{kallmann2014dynamic}, but since the total
number of additional triangles created during refinement is small
(about 5\% in our experiments), their impact on the overall cost of
path planning remains negligible.

In order to use our roadmap synthesis method for actual path planning
applications, additional problems need to be solved, which have not
been addressed in this article. First, the roadmap graph usually needs
to be connected to specific origin and destination locations. We perform this
operation by combining solutions proposed in~\cite{jan2014shortest}
and~\cite{kallmann2014dynamic}. In a few words, the technique consists
in triangulating the set of obstacles after adding the origin and
destination locations as additional vertices, then removing them, and
finally performing some local operations for taking care of obstacles
that are close to these locations. A second problem is to search for a
suitable path in a roadmap graph, which can be achieved by
shortest-path search algorithms such as Dijkstra's or A$^*$, using a
metric of interest.

Finally, it is worth mentioning that our motivation for studying
roadmap graph generation was prompted by our participation to an
autonomous robotics contest\footnote{\texttt{http://www.eurobot.org}},
in which mobile robots compete in a $3 \times 2~$m$^2$ area heavily
constrained with obstacles. In this setting, achieving very small
computation times was essential in order to be competitive.

\bibliographystyle{plain}
\bibliography{biblio}

\begin{thebibliography}{10}

\bibitem{ali2013modeling}
Saad Ali, Ko~Nishino, Dinesh Manocha, and Mubarak Shah.
\newblock {\em Modeling, Simulation and Visual Analysis of Crowds: A
  Multidisciplinary Perspective}.
\newblock Springer, 2013.

\bibitem{bern2004triangulations}
Marshall Bern.
\newblock Triangulations and mesh generation.
\newblock In Jacob~E Goodman and Joseph O'Rourke, editors, {\em Handbook of
  Discrete and Computational Geometry}, pages 563--582. Chapman \& Hall/CRC,
  2nd edition, 2004.

\bibitem{bhattacharya2008roadmap}
Priyadarshi Bhattacharya and Marina~L. Gavrilova.
\newblock Roadmap-based path planning-using the {V}oronoi diagram for a
  clearance-based shortest path.
\newblock {\em Robotics \& Automation Magazine, IEEE}, 15(2):58--66, 2008.

\bibitem{chew1989constrained}
L.~Paul Chew.
\newblock Constrained {D}elaunay triangulations.
\newblock {\em Algorithmica}, 4:97--108, 1989.

\bibitem{berg2008computational}
Mark de~Berg, Otfried Cheong, Marc van Kreveld, and Mark~H. Overmars.
\newblock {\em Computational Geometry: Algorithms and Applications}.
\newblock Springer-Verlag TELOS, 3rd edition, 2008.

\bibitem{demyen2006efficient}
Douglas Demyen and Michael Buro.
\newblock Efficient triangulation-based pathfinding.
\newblock In {\em Proceedings of the 21st national conference on Artificial
  intelligence}, pages 942--947. AAAI Press, 2006.

\bibitem{devillers1998improved}
Olivier Devillers.
\newblock Improved incremental randomized {D}elaunay triangulation.
\newblock In {\em Proceedings of the 14th annual symposium on Computational
  geometry}, pages 106--115. ACM, 1998.

\bibitem{ge2002dynamic}
Shuzhi~S. Ge and Yun~J. Cui.
\newblock Dynamic motion planning for mobile robots using potential field
  method.
\newblock {\em Autonomous Robots}, 13(3):207--222, 2002.

\bibitem{geraerts2010planning}
Roland Geraerts.
\newblock Planning short paths with clearance using explicit corridors.
\newblock In {\em ICRA'10: Proceedings of the IEEE International Conference on
  Robotics and Automation}, pages 1997--2004. IEEE, 2010.

\bibitem{jan2014shortest}
Gene~Eu Jan, Chi-Chia Sun, Wei~Chun Tsai, and Ting-Hsiang Lin.
\newblock An $\mathcal{O}(n\log n)$ shortest path algorithm based on {D}elaunay
  triangulation.
\newblock {\em Mechatronics, IEEE/ASME Transactions on}, 19(2):660--666, 2014.

\bibitem{kallmann2010shortest}
Marcelo Kallmann.
\newblock Shortest paths with arbitrary clearance from navigation meshes.
\newblock In {\em Proceedings of the 2010 ACM SIGGRAPH/Eurographics Symposium
  on Computer Animation (SCA)}, pages 159--168. Eurographics Association, 2010.

\bibitem{kallmann2014dynamic}
Marcelo Kallmann.
\newblock Dynamic and robust local clearance triangulations.
\newblock {\em ACM Transactions on Graphics}, 33(5):161:1--161:17, 2014.

\bibitem{kambhampati1985multiresolution}
Subbarao Kambhampati and Larry~S. Davis.
\newblock Multiresolution path planning for mobile robots.
\newblock Technical report, DTIC Document, 1985.

\bibitem{lamarche2004crowd}
Fabrice Lamarche and St{\'e}phane Donikian.
\newblock Crowd of virtual humans: {A} new approach for real time navigation in
  complex and structured environments.
\newblock {\em Computer Graphics Forum}, 23(3):509--518, 2004.

\bibitem{latombe1991planning}
Jean-Claude Latombe.
\newblock {\em Robot Motion Planning}.
\newblock Kluwer Academic Publishers, 1991.

\bibitem{lavalle2006planning}
Steven~M. LaValle.
\newblock {\em Planning algorithms}.
\newblock Cambridge University Press, 2006.

\bibitem{lavalle01}
Steven~M. LaValle and James J.~Kuffner Jr.
\newblock Randomized kinodynamic planning.
\newblock {\em I. J. Robotic Res.}, 20(5):378--400, 2001.

\bibitem{wagner2007speed}
Dorothea Wagner and Thomas Willhalm.
\newblock Speed-up techniques for shortest-path computations.
\newblock In {\em STACS 2007}, pages 23--36. Springer, 2007.

\bibitem{wein2005visibility}
Ron Wein, Jur~P. van~den Berg, and Dan Halperin.
\newblock The visibility--{V}oronoi complex and its applications.
\newblock In {\em Proceedings of the twenty-first annual symposium on
  Computational geometry}, pages 63--72. ACM, 2005.

\end{thebibliography}

\end{document}